\documentclass[prl, twocolumn]{revtex4}

\usepackage{epsfig}
\usepackage{graphicx}
\usepackage{amsmath}
\usepackage{amsfonts}
\usepackage{epstopdf}
\usepackage{array}
\usepackage{hyperref}
\usepackage{txfonts}

\renewcommand{\vec}{\boldsymbol}
\newcommand{\rmi}{\mathrm{i}}
\newcommand{\rme}{\mathrm{e}}
\newcommand{\rmd}{\mathrm{d}}

\begin{document}

\title{Phase diagram of the rotating two-component Fermi gas including vortices}

\author{Harmen J. Warringa}
\email{warringa@th.physik.uni-frankfurt.de}

\affiliation{Institut f\"ur Theoretische Physik, Goethe-Universit\"at
  Frankfurt am Main, Germany\\
and Frankfurt Institute for Advanced Studies (FIAS), Frankfurt am Main, Germany}

\newcommand{\sumint}{\sum \!\!\!\!\!\!\!\! \int \;}

\date{\today}

\begin{abstract}
  We determine the conditions under which superfluidity with and
  without quantized vortices appears in a weakly interacting
  two-component atomic Fermi gas that is trapped in a rotating
  cylindrical symmetric harmonic potential. We compute the phase
  diagram as a function of rotation frequency, scattering length,
  temperature, total number of trapped atoms, and population
  imbalance.
\end{abstract}

\maketitle

{\it Introduction.}  Superfluids are fluids that can flow with hardly
any friction. If a superfluid is put into a rotating container, vortices
carrying quantized circulation can be formed. These vortices are the
hallmark of superfluidity and have been observed experimentally in
Helium-4, in atomic Bose-Einstein condensates and in a two-component
atomic Fermi gas made out of $^6$Li atoms \cite{Zwierlein05,
  Zwierlein06}.

In this Letter we will focus on vortex formation in an equilibrated
weakly interacting two-component Fermi gas that is trapped in a
rotating cylindrical symmetric harmonic potential. The two components
of this gas consist of atoms in different hyperfine states, and will
be labeled by $\uparrow, \downarrow$. We will consider a situation in
which both components have equal mass $M$. In experiment, one can vary
the strength of the interaction between the components using the
Feshbach resonance.  Furthermore one can control the rotation
frequency $\Omega$, temperature $T$, total number of atoms, and the
relative difference between the number of atoms in each component $P$
(population imbalance) \cite{Zwierlein05, Zwierlein06}.  When the
interaction is tuned to be attractive, the gas is cooled to low enough
$T$, and $\Omega$ and $P$ are made small enough, the components will
form Cooper pairs resulting in a Bardeen-Cooper-Schrieffer (BCS)
superfluid \cite{Stoof96}.

The aim of this Letter is to determine in which region of the
parameter space that is accessible to experiment vortices will be
formed. So far only very limited information about this region is
available. Only for weak interactions, $T=0$ and $P=0$, the lower
critical $\Omega$ has been obtained theoretically. This frequency was
estimated in \cite{Bruun01} using a Ginzburg-Landau approach and
computed by the author through solving the Bogoliubov-de Gennes (BdG)
equation \cite{Warringa11}. Experimentally only the critical $P$ for
vortex formation at one specific $\Omega$ has been determined
for different scattering lengths in the strongly interacting regime
\cite{Zwierlein06}.

We will consider the following trapping potential: $U(\vec x) =
\frac{1}{2} M \omega^2 \rho^2$, where we have introduced cylindrical
coordinates, i.e. $\vec x = ( \rho \cos \phi, \rho \sin \phi, z)$. We
will allow the trap to rotate with constant angular frequency $\Omega$
in the $x$-$y$ plane. This trapping potential implies harmonic
confinement with frequency $\omega$ in the $x$-$y$ plane and infinite
extension in the $z$-direction. In an experiment this situation can be
approached by choosing the trapping frequency in the $z$-direction
($\omega_z$) much smaller than $\omega$. The characteristic length
scale of the potential is the harmonic oscillator length $\lambda =
\sqrt{\hbar / M \omega}$.  In the setup of Ref.~\cite{Zwierlein06} the
radial trapping frequency was taken to be $\omega / (2\pi) = 110\;
\mathrm{Hz}$. In that case $\lambda \sim 3.9\;\mathrm{\mu m}$ and the
characteristic energy scale $\hbar \omega / k_{\mathrm B} \sim 5.3\;
\mathrm{nK}$.

The order parameter for superfluidity is the pairing field
$\Delta(\vec x)$.  For a single vortex that is located at the center
of the trap it has the following form: $\Delta(\vec x) = \tilde
\Delta(\rho) \exp( \rmi k \phi)$, with $k$ the winding number of the
vortex. The $k=0$ case corresponds to a superfluid without
vortices. We will assume that $\tilde \Delta(\rho) \in
\mathbb{R}$. There is no superfluidity at the core of a vortex, hence
$\tilde \Delta(\rho = 0) = 0$ for $k \neq 0$.  If $\Omega=0$, $T=0$,
and $P=0$, the whole system of atoms forms a vortex-free
superfluid. In that situation the vortex is metastable. This is
because condensation energy is lost near the vortex core, the atoms
traveling around the vortex core have an average velocity resulting in
a kinetic energy cost, and the system has to expand to compensate for
the density depletion at the vortex core. This expansion costs energy
because of the attractive interaction.

Starting from a situation in which $\Omega =0$, let us now imagine
increasing $\Omega$. If both $T=0$ and $P=0$, the whole system stays
superfluid up to a critical frequency $\Omega_b$. Above $\Omega_b$
part of the system will turn into a normal gas due to breaking of
Cooper pairs~\cite{Pairbreaking}. If either $T \neq 0$ or $P \neq 0$
there are unpaired atoms present for any nonzero $\Omega$. The
unpaired atoms are predominantly located in the outer regions of the
system. They rotate like a rigid body and acquire therefore rotational
energy. A vortex is also a source of rotational energy since it
induces angular momentum in the system.  If the rotational energy
gains overcome the costs, a single vortex will be preferred above a
frequency $\Omega_l$. Because of symmetry and energy arguments, this
vortex has unit winding number ($k=1$) and is located at the center of
the trap ($\rho=0$). By further increasing $\Omega$ more vortices can
be created resulting in a vortex lattice \cite{Feder04, Tonini08}. At
the same time increasing $\Omega$ will shrink the size of the
superfluid region \cite{Zhai06}, while the system as a whole will
expand due to the centrifugal force. At some point the superfluid
region will be so small that it can only support a single vortex at
$\rho = 0$. By further increasing $\Omega$ this vortex will disappear
at a upper critical rotation frequency $\Omega_u$. Then at an even
larger rotation frequency $\Omega_s$ superfluidity will vanish
completely via a second order transition \cite{Feder04}. The upper
critical frequency for superfluidity $\Omega_s$ has been computed in
Refs.~\cite{Zhai06, Veillette06} for a balanced gas using the local
density approximation (see also \cite{uppercrit} for related analyses
of the phase boundary of superfluidity).  In this Letter we will
obtain $\Omega_s$ by solving the BdG equation.  Finally, above $\Omega
= \omega$ the system will be torn apart.

As pointed out above, the first vortex that appears and the last
vortex that disappears when increasing $\Omega$ has $k=1$ and is
located at $\rho=0$. Moreover, in experiment it has been observed that
the number of vortices goes continuously to zero when increasing $P$
at fixed $\Omega$ \cite{Zwierlein06}. Therefore, we can map out the
entire region in which one or more vortices are preferred
thermodynamically by determining the conditions under which the $k=1$
situation has lower Helmholtz free energy than the $k=0$
situation. For studies of real-time dynamics of vortex formation we
refer to Refs.~\cite{Tonini08, Bulgac11}.  To obtain $\Omega_s$ we
will determine the point at which $\Delta(\rho = 0)$ vanishes for
$k=0$.

{\it Setup.} We will now briefly discuss the details of the
calculation, a more extensive discussion can be found in our earlier
work~\cite{Warringa11}. To obtain the Helmholtz free energy one first
needs to compute the pairing field $\Delta(\vec x) = \tilde
\Delta(\rho) \exp( \rmi k \phi)$ and the density profiles
$\rho_{\uparrow \downarrow}(\vec x) = \rho_{\uparrow
  \downarrow}(\rho)$ for each component separately. For a weakly
interacting gas these can be found by solving the BdG
equation self-consistently:
\begin{equation}
\left(
\begin{array}{cc}
 \mathcal{H}_\uparrow(\Omega)
&
\Delta(\vec x )
\\
\Delta^*(\vec x)
&
 -\mathcal{H}^*_\downarrow(\Omega)
\end{array}
\right)
\left(
\begin{array}{c}
u_i(\vec x)
\\
v_i(\vec x)
\end{array}
\right)
= 
E_i
\left(
\begin{array}{c}
u_i(\vec x)
\\
v_i(\vec x)
\end{array}
\right).
\label{eq:hbdg}
\end{equation}
Here $\mathcal{H}_{\uparrow \downarrow}$ contains the single-particle
Hamiltonian, the Hartree self-energy, and the chemical potential
$\mu_{\uparrow \downarrow}$. Explicitly, it reads $
\mathcal{H}_{\uparrow, \downarrow}(\Omega) = \frac{\vec p^2}{2M} +
\frac{1}{2} M \omega^2 \rho^2 - \Omega L_z - \mu_{\uparrow,
  \downarrow} \! +\! g n_{\downarrow, \uparrow}(\rho), $ where the
$z$-component of the angular momentum is given by $L_z = -\rmi
\hbar \partial/\partial \phi$, and $g = 4 \pi a \hbar^2 / M$ is the
coupling constant with $a$ the $s$-wave scattering length. The
wavefunctions $u_i(\vec x)$ and $v_i(\vec x)$ have to be normalized as
$\int \rmd^3 x \, \left[ \vert u_i(\vec x) \vert^2 + \vert v_i(\vec x
  ) \vert^2 \right] = 1$. The number densities are given by
$n_{\uparrow}(\vec x) = \sum_i f(E_i) \vert u_i(\vec x) \vert^2$ and
$n_{\downarrow}(\vec x) = \sum_i f(-E_i) \vert v_i(\vec x) \vert^2$,
where $f(E) = [\exp(\beta E) + 1]^{-1}$ with $\beta = 1/(k_\mathrm{B}
T)$. The pairing field follows from the regular (reg) part of the
anomalous propagator in the following way $\Delta(\vec x) = g
G^{\mathrm{reg}}_{\uparrow \downarrow}(\vec x, \vec x) $ where
$G_{\uparrow \downarrow}(\vec x, \vec x') = \sum_i f(E_i) u_i(\vec x)
v_i^*(\vec x')$. To obtain this regular part we have used a method
\cite{Warringa11} based on the procedures discussed in
Refs.~\cite{Bruun99, BulGras}.

We will consider a fixed number of atoms per unit length $\lambda$ in
the $z$-direction, and denote this number by $\mathcal{N}_{\uparrow,
  \downarrow} = \lambda/L \int \rmd^3 x\, n_{\uparrow,
  \downarrow}(\vec x)$. Here $L$ is the length of the system in the
$z$-direction which we will take to be infinite.  We will write
$\mathcal{N} = \mathcal{N}_{\uparrow} + \mathcal{N}_{\downarrow}$ for
the total number of atoms per unit length. The population imbalance or
polarization is defined as $P = ( \mathcal{N}_{\uparrow}-
\mathcal{N_\downarrow}) / \mathcal{N}$.  The chemical potentials
$\mu_{\uparrow, \downarrow}$ will be solved for such that the required
$\mathcal{N}_{\uparrow,\downarrow}$ is obtained.

To solve the BdG equation numerically, we have discretized the radial
part of the wavefunctions on a Lagrange mesh \cite{DVR} based on
Maxwell polynomials \cite{Warringa11}. Typically we could reach a
relative accuracy of order $10^{-3}$ with about $64$ to $96$ mesh
points for $\mathcal{N} = 1000$. The angular and $z$-dependence of the
wavefunctions were treated exactly. Integration over the $z$-momentum
was performed using the adaptive Simpson method. To solve for
self-consistency we have used the Newton-Broyden rootfinding
method. The full details of our numerical procedure are explained in
\cite{Warringa11}. Examples of pairing field and density profiles with
and without a vortex can be found in Refs.~\cite{Bruun99, profiles,
  Machida06, Warringa11}.  In particular we have obtained excellent
agreement with Ref.~\cite{Machida06}, in which a vortex profile for an
imbalanced gas that is also trapped in a cylindrical symmetric harmonic
potential is presented.

Once the pairing field and the density profiles have been obtained,
the Helmholtz free energy per unit length (which is ultraviolet
finite) can be computed in the following way
\begin{multline}
\mathcal{F}
= 
- 
\frac{\lambda}{L}
\sum_{i} \left[ \frac{ \vert E_i \vert}{2}   
+ \frac{1}{\beta} \log \left(1+\rme^{-\beta \vert E_i \vert} \right)
\right]
+ \mu_\uparrow \mathcal{N}_\uparrow + \mu_\downarrow \mathcal{N}_\downarrow
\\
- 
\frac{\lambda}{L}
\int \rmd^3 x \,
\left[
G_{\uparrow \downarrow}(\vec x, \vec x)^*
\Delta(\vec x)
+
g 
n_{\uparrow}(\vec x) n_{\downarrow}(\vec x ) \right]
+ 
\frac{\lambda}{L}
\sum_{i} \epsilon_i,
\label{eq:helmholtz}
\end{multline}
where $\epsilon_i$ are the eigenvalues of the Hartree-Fock Hamiltonian
$H_{\mathrm{HF}} = \left[\mathcal{H}_{\uparrow}(\Omega = 0) +
  \mathcal{H}_{\downarrow}(\Omega = 0)\right]/2$. We have computed
$\Delta \mathcal{F} = \mathcal{F}_{k=1} - \mathcal{F}_{k=0}$ for
several values of the external parameters and obtained the phase
boundary by determining the point at which $\Delta \mathcal{F} =
0$. Typically we could locate this boundary with a relative accuracy
of about $10^{-2}$ to $10^{-3}$.

{\it Results.}  We will now present several phase diagrams from which
it can be seen for which values of the rotation frequency $\Omega$,
scattering length $a$, temperature $T$, total number of atoms per unit
length $\mathcal{N}$, and imbalance $P$, superfluidity with one or
more vortices will be formed (light gray area). Also we will indicate
in these phase diagrams when the system exhibits superfluidity without
any vortex (dark gray area). As a measure of the interaction strength
we will use $1/\vert k_{F0} a\vert$, where $k_{F0}$ denotes the Fermi
momentum of the superfluid at $T=0$, $P=0$, $\Omega=0$ and $\rho =0$.

\begin{figure}[t]
\includegraphics[scale=0.9]{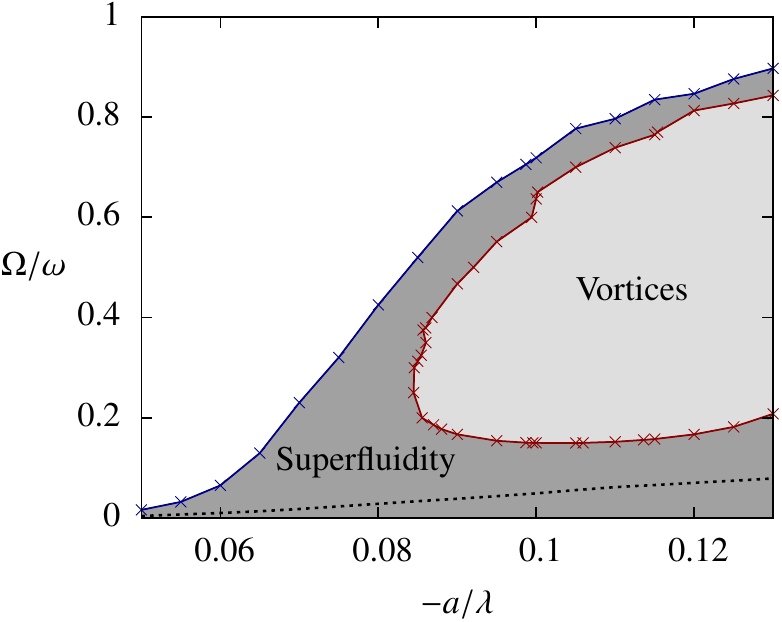}
\caption{Phase diagram: $a$--$\Omega$ plane, 
  for $T \approx 0$, $\mathcal{N} = 1000$ and $P = 0$. 
  The dotted line indicates $\Omega_b$.
  Here $0.8 \leq 1/(k_{F0}\vert a \vert) \leq 2.6$.
\label{fig:oa}
}
\end{figure}

\begin{figure}[t]
\includegraphics[scale=.8]{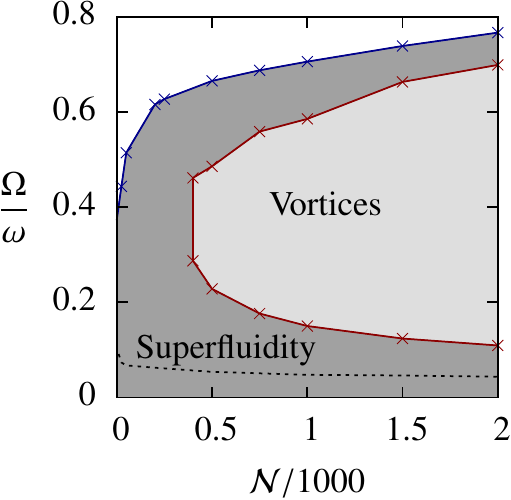}
\includegraphics[scale=.8]{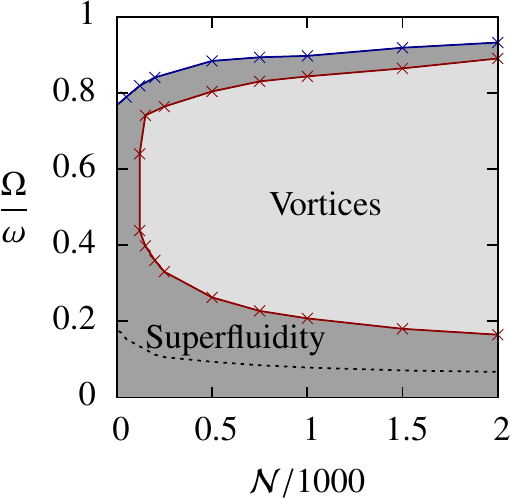}
\caption{Phase diagrams: $\mathcal{N}$--$\Omega$ plane, for $T\approx0$, 
$P=0$, and
 $1/(k_{F0}\vert a \vert) = 1.2$ (left) and $1/(k_{F0}\vert a \vert) = 0.8$ (right).
\label{fig:ol}
}
\end{figure}

In Ref.~\cite{Warringa11}, we have computed the critical frequency for
unpairing ($\Omega_b$) and the lower critical frequency for vortex
formation ($\Omega_l$) as a function of scattering length for $T=0$,
$P=0$ and $\mathcal{N}=1000$. In Fig.~\ref{fig:oa} we display the full
phase diagram at $T\approx 0$. Here we write $T\approx 0$ to indicate
that the vortex phase boundaries and the unpairing transition were
computed at $T=0$ exactly, whereas $\Omega_s$ was computed at $T=0.01
\hbar \omega$. We have used this very small $T$ to ensure that $\tilde
\Delta(\rho = 0)$ approaches zero continuously when increasing
$\Omega$, so that we could determine $\Omega_s$.  Exactly at $T=0$
$\tilde \Delta(\rho= 0)$ does not seem to vanish when increasing
$\Omega$, although it does become very small.

The successive transitions that one encounters when increasing
$\Omega$ were already described in the introduction and can be clearly
seen in the diagram. The first transition is the unpairing transition
occurring at $\Omega=\Omega_b$. It is of second order and turns into a
cross-over for $T>0$. Therefore it is a quantum phase transition and
at $\Omega_b$ the system resides at a quantum critical point.  If
$\vert a \vert$ increases, it will become more difficult to break the
Cooper pairs, hence $\Omega_b$ grows in that case. For $\vert a\vert
\gtrsim 0.085 \lambda$ vortices will be formed for $\Omega_l \leq
\Omega \leq \Omega_u$.  The structure of $\Omega_l$ is a result of the
interplay of two effects \cite{Warringa11}. The first is that the
energy cost of a vortex at $\Omega=0$ increases when increasing $\vert
a \vert$. That naturally leads to a larger $\Omega_l$. The second
effect is that unpairing becomes easier for smaller $\vert a \vert$,
hence more rotational energy can be acquired by the $k=0$
superfluid. This leads to an increase of $\Omega_l$ for weak
interactions and is the reason that for small $\vert a\vert$ no
vortices will be formed for any $\Omega$.  The size of the superfluid
region shrinks above $\Omega_b$ when increasing $\Omega$. If $\vert
a\vert$ grows at fixed $\Omega$, it becomes more difficult to destroy
superfluidity by rotation. This results in a larger superfluid region,
so that a vortex can fit more easily. For these reasons both
$\Omega_l$ and the upper critical frequency for superfluidity
$\Omega_s$ grow with increasing $\vert a \vert$.  The behavior of
$\Omega_s$ is qualitatively in agreement with the results of
Refs.~\cite{Zhai06, Veillette06}.

This phase diagram will be modified quantitatively when changing
$\mathcal{N}$. However, the qualitative structure will remain the same
if $\mathcal{N}$ is large enough. To illustrate this, we display in
Fig.~\ref{fig:ol} the phase diagram in the $\mathcal{N}$-$\Omega$
plane for two different interaction strengths. A larger $\mathcal{N}$
implies a larger system, and hence for a given $\Omega$ the atoms at
the boundaries have a larger velocity. This makes pairing more
difficult, leading to a smaller $\Omega_b$ as can be seen in
Fig.~\ref{fig:ol}.  To explain the behavior of $\Omega_l$, we can use
that the rotational energy gain of the vortex is proportional to the
angular momentum which is proportional to $\mathcal{N}$. The energy
costs of the vortex grow much slower when increasing
$\mathcal{N}$. Hence a larger $\mathcal{N}$ leads to a smaller
$\Omega_l$. Below a certain $\mathcal{N}$ no vortices will be formed
for any $\Omega$. The upper critical frequencies $\Omega_u$ and
$\Omega_s$ increase if $\mathcal{N}$ grows.

\begin{figure}[t]
\includegraphics[scale=.8]{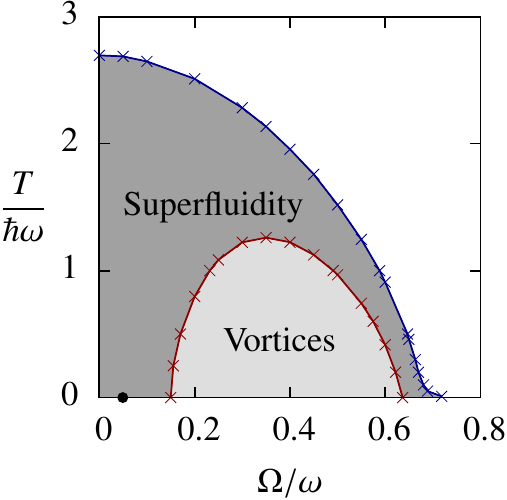}
\includegraphics[scale=.8]{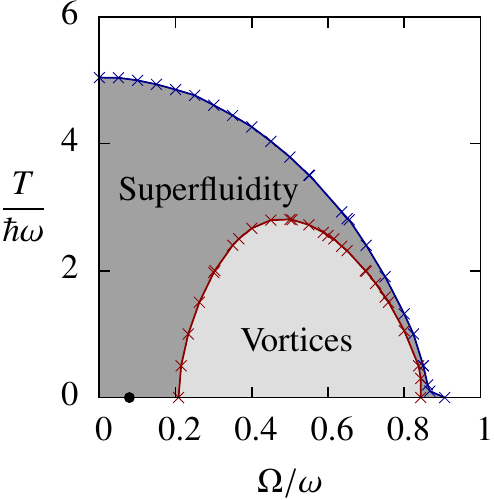}
\caption{Phase diagrams: $\Omega$--$T$ plane, for $\mathcal{N} =
  1000$, $P = 0$ and $a = -0.10 \lambda$ (left, $1/(k_{F0}\vert
  a \vert) = 1.2$) and $a = -0.13 \lambda$ (right,
  $1/(k_{F0}\vert a \vert) = 0.8$). The dot indicates the unpairing
  quantum critical point.
\label{fig:ot}
}
\end{figure}

If $T$ is increased the window in $\Omega$ in which vortices are
formed narrows. This can be seen from Fig.~\ref{fig:ot} in which we
display the phase diagram in the $\Omega$-$T$ plane for
$\mathcal{N}=1000$, $P=0$ and two different scattering lengths. Above
a certain $T$, vortices will not be formed for any $\Omega$ while the
system is still partly superfluid. It is the easiest to make vortices
at intermediate $\Omega$, since in that case the window in $T$ is the
largest. This $\Omega$-$T$ phase diagram is qualitatively very similar
to that of a rotating Bose-Einstein condensate \cite{Stringari99}.

\begin{figure}[t]
\includegraphics[scale=.8]{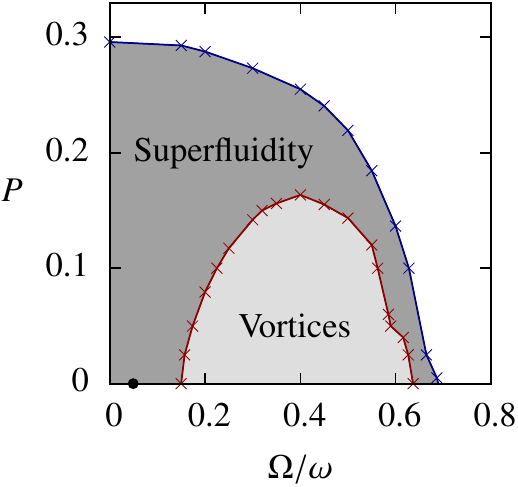}
\includegraphics[scale=.8]{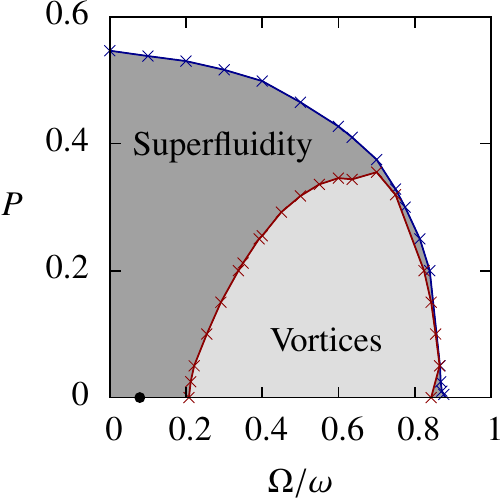}
\caption{As in Fig.~\ref{fig:ot}, but now in $\Omega$--$P$ plane for $T=0$.
\label{fig:op}
}
\end{figure}

In Fig.~\ref{fig:op} we display the phase diagram in the $\Omega$-$P$
plane for $T=0$ and two different scattering lengths. The larger the
$P$, the narrower the window in $\Omega$ in which vortices are formed
becomes. Above a certain $P$ part of the system can still be
superfluid, but vortices will not be formed for any $\Omega$.

From Figs.~\ref{fig:ot} and \ref{fig:op} it can be seen that the
critical $T$ and critical $P$ for vortex formation and superfluidity
grow when increasing the interaction strength. When $\Omega$ is
increased, $\Omega_s$ always decreases. Especially for large
interaction strengths $\Omega_u$ and $\Omega_s$ lie very close each
other.

\begin{figure}[t]
\includegraphics[scale=.8]{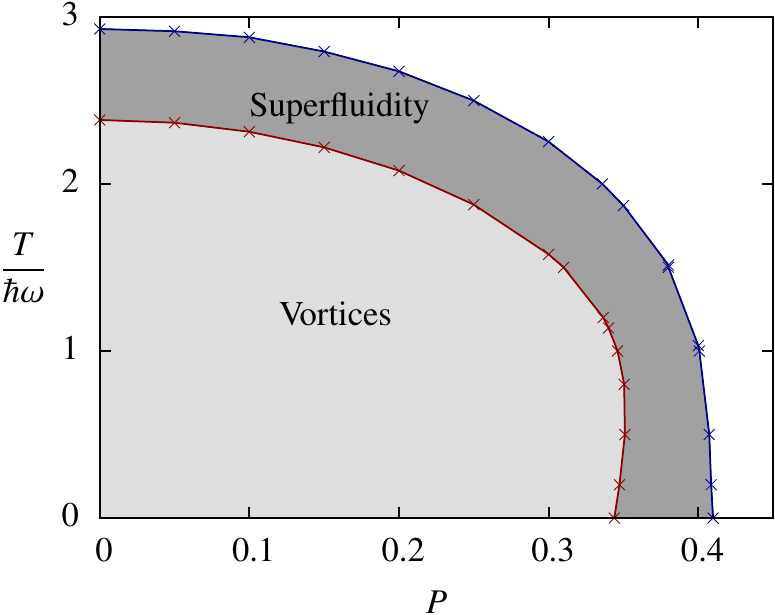}
\caption{Phase diagram: $P$-$T$ plane, 
for $\mathcal{N} = 1000$, $\Omega=0.636 \omega$, and $a =
  -0.13 \lambda$. Here $1/(k_{F0}\vert a \vert) = 0.8$.
\label{fig:pt}
}
\end{figure}

\begin{figure}[t]
\includegraphics[scale=.8]{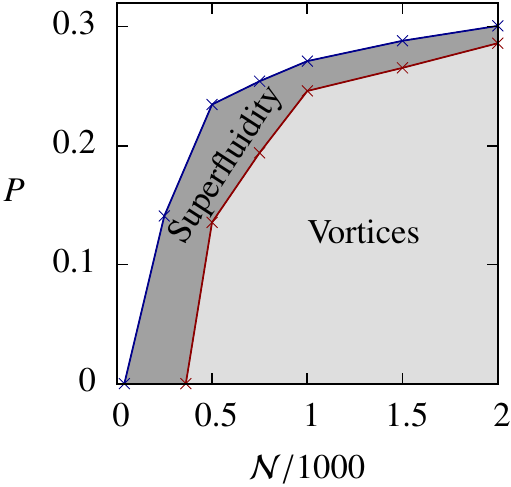}
\includegraphics[scale=.8]{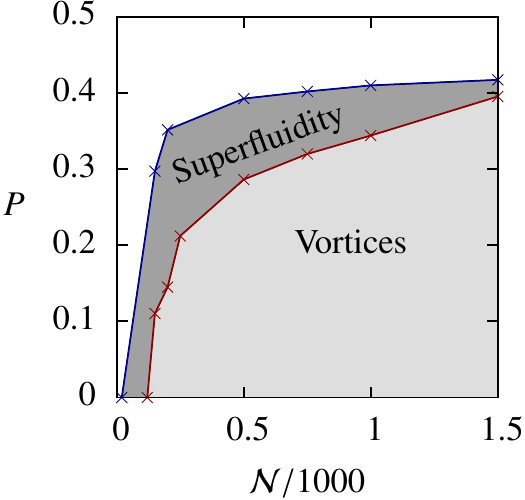}
\caption{Phase diagrams: $\mathcal{N}$-$P$ plane, 
for $T=0$, $\Omega = 0.636 \omega$, and
 $1/(k_{F0}\vert a \vert) = 1.0$ (left)
and $1/(k_{F0}\vert a \vert) = 0.8$ (right).
\label{fig:pc}
}
\end{figure}

In the experiment described in Ref.~\cite{Zwierlein06} a strongly
interacting two-component Fermi gas made out of $\sim 7 \times 10^6$
atoms was trapped in a cigar-shaped potential with a frequency ratio
of $\omega_z/\omega = 23/110$. Using the Thomas-Fermi approximation at
$T=0$ we estimate that in this experiment $\mathcal{N}(z=0) \sim 1
\times 10^5$. The atoms were stirred with a laser at a frequency of
$\Omega = (70/110) \omega \approx 0.636 \omega$. After stirring, one
waited until the system had equilibrated and measured the number of
vortices as a function of $P$. In this way an upper critical imbalance
for vortex formation ($P_c$) could be determined. Although this
situation is not completely equivalent to our setup it is nevertheless
interesting to make a comparison to this experiment. Therefore we
display in Fig.~\ref{fig:pt} the phase diagram in the $P$-$T$ plane
for $\Omega = 0.636 \omega$, $\mathcal{N}=1000$, and $1/(k_{F0}\vert a
\vert) = 0.8$.  In qualitative agreement with the experiment it can be
seen that $P_c$ is weakly dependent on $T$ for small temperatures. We
find that for $P>P_c$ there is a small window in which part of the
system is in the superfluid phase without any vortex, as in the
experiment.

In Fig.~\ref{fig:pc} we display the phase diagram in the
$\mathcal{N}$-$P$ plane for $T=0$, $\Omega = 0.636 \omega$ and two
different interaction strengths. As expected and seen in experiment,
weaker interactions imply a lower $P_c$.  In the experiment it was
found that $P_c \sim 0.1$ at $1/(k_{F0}\vert a \vert) =
0.5$~\cite{Zwierlein06}.  The values of $P_c$ we have obtained seem to
be large compared to this value, because we consider weaker
interactions and our $\mathcal{N}(z=0)$ is much
smaller. This quantitative difference could have been caused by the
different stirring method or by the shape of the trapping
potential. Also it could signal that one should take into account
beyond the mean field corrections at $1/(k_{F0}\vert a \vert) = 0.8$.

{\it Conclusions.}  For the first time we have unveiled the full phase
diagram of a trapped, rotating, and weakly-interacting two-component
Fermi gas including vortices.  We have made detailed predictions for
the conditions under which superfluidity with and without vortices is
formed as a function of rotation frequency, scattering length,
temperature, number of atoms and population imbalance. The phase
diagrams we have obtained are quantitatively reliable and are in
principle directly comparable to a possible future experimental
determination. Our analysis can be extended to more complicated
systems, like Fermi gases with $p$-wave pairing, Fermi gases with
more than two components, and Fermi gases in which the two components
have unequal mass. This will be useful for the experimental search for
superfluidity in such systems.

{\it Acknowledgments.}  The work of H.J.W.\ was supported by the
Extreme Matter Institute (EMMI). I would like to thank Armen
Sedrakian, Henk Stoof, Massimo Mannarelli, Lianyi He, Xu-Guang Huang,
and Dirk Rischke for useful discussions. I am grateful to the Center
for Scientific Computing Frankfurt for providing computational
resources.

\end{document}